\documentclass{eptcs}

\usepackage{paper}

\newsavebox{\tempboxa}

\def\doframeit#1{\vbox{%
  \hrule height\fboxrule
    \hbox{%
      \vrule width\fboxrule \kern\fboxsep
      \vbox{\kern\fboxvsep #1\kern\fboxvsep }%
      \kern\fboxsep \vrule width\fboxrule }%
    \hrule height\fboxrule }}
    
\def\frameit{\smallskip \advance \linewidth by -7.5pt \setbox0=\vbox
\bgroup
\strut \ignorespaces }

\def\endframeit{\ifhmode \par \nointerlineskip \fi \egroup
\doframeit{\box0}}

\newdimen \fboxvsep
\fboxvsep=\fboxsep
\advance \fboxsep by -8pt

\begin{document}

\title{A Case for Dynamic Reverse-code Generation \\to Debug Non-deterministic Programs}

\author{Jooyong Yi
\institute{School of Computing \\ National University of Singapore}
\email{jooyong@comp.nus.edu.sg}
}
\def\titlerunning{A case for dynamic reverse-code generation}
\def\authorrunning{J. Yi}

\maketitle

\begin{abstract}
Backtracking (i.e., reverse execution) helps the user of a debugger to naturally think backwards along the execution path of a program, and thinking backwards makes it easy to locate the origin of a bug. 
So far backtracking has been implemented mostly by state saving or by checkpointing. These implementations, however, inherently do not scale. 
Meanwhile, a more recent backtracking method based on reverse-code generation seems promising because executing reverse code can restore the previous states of a program without state saving.
In the literature, there can be found two methods that generate reverse code: (a) static reverse-code generation that pre-generates reverse code through static analysis before starting a debugging session, and (b) dynamic reverse-code generation that generates reverse code by applying dynamic analysis on the fly during a debugging session. 
In particular, we espoused the latter one in our previous work to accommodate non-determinism of a program caused by e.g., multi-threading. 
To demonstrate the usefulness of our dynamic reverse-code generation, this article presents a case study of various backtracking methods including ours.
We compare the memory usage of various backtracking methods in a simple but nontrivial example, a bounded-buffer program. 
In the case of non-deterministic programs such as this bounded-buffer program, our dynamic reverse-code generation  outperforms the existing backtracking methods in terms of memory efficiency.
\end{abstract}

\section{Introduction} \label{sec:intro}

When something goes wrong, we people tend to think backwards to find the cause of an observed problem.
As a simple example, if you lost something valuable, you would go back to the places you visited during the day, one by one starting from the last place you visited. 
When a program goes wrong, it is tempting to do the same for debugging. 
One would wish to trace backwards an erroneous execution path to find the cause of unexpected behavior of a program.

However, alas, programs run only forwards. 
As a result, traditional debugging methods are by and large speculation-based. 
Developers first guess program points that are potentially problematic, and observe program states at those program points by inserting print commands or using breakpoints. 
Such speculation-based debugging is often time-consuming, and requires high expertise in the program being debugged.

Indeed, empirical studies of Ko and Myers~\cite{Ko-Myers-CHI09, Ko-Myers-ICSE08} suggest that backward reasoning, which is supported through their custom debugger, makes a debugging process more efficient as compared to the traditional speculation-based methods; 
in their user experiments, users could complete more debugging tasks successfully in a shorter time when backward reasoning is facilitated than when a traditional breakpoint-based method is used. 

In fact, the idea of supporting backtracking in debugging is not new. 
There have been numerous studies that make use of backtracking for debugging~\cite{Agrawal-DeMillo-Spafford-IEEE91,Agrawal-DeMillo-Spafford-SPE93,Akgul-Mooney-ACM04,Balzer-AFIPS69,Biswas-Mall-SIGPLAN99,Booth-Jones-AADEBUG97,Boothe-PLDI00,Cook-CJ02,Ko-Myers-CHI09, Ko-Myers-ICSE08}. 
However, all the backtracking methods used in those studies share a common problem; 
there is a limit in the length of an execution path that can be traced backwards.
Program states beyond that limit cannot be restored using those backtracking methods. 
This is because all backtracking methods rely on state saving or checkpointing (\ie periodic state saving) where changes of program states are saved in the memory.
Thus, as a program runs longer, more memory is consumed in general.

However, one backtracking method does not entirely depend on state saving. 
One can restore the previous program states by running special separate code. 
We call such special code \emph{reverse code}.
As a simple example, an assignment command \stt{x:=x+1} can be traced backward by executing reverse code \stt{x:=x-1}.
Such reverse code can be generated through program analysis as evidenced by Akgul and Mooney~\cite{Akgul-Mooney-ACM04} and in our previous work~\cite{Lee-VD06}.
As one can imagine, it is impossible to obtain such reverse code for every command that is executed.
Only in such cases, state saving is used.
Such less dependence on state saving results in more applicability of backtracking.

Reverse code can be generated through static analysis~\cite{Akgul-Mooney-ACM04} or dynamic analysis~\cite{Lee-VD06}.
We refer to these two methods as static reverse-code generation and dynamic reverse-code generation, respectively.
The former method prepares reverse code in advance before starting a debugging session, whereas the latter method generates reverse code, which works only for the current execution path, on the fly during a debugging session. 
While the difference between these two methods will be detailed in Section~\ref{sec:bm}, we first point out that non-determinism caused by multi-threading has a different impact on each method. 
Non-determinism makes almost no impact on dynamic reverse-code generation;
reverse code can be generated as long as the current execution path can be obtained.
Meanwhile, it is challenging to accommodate non-determinism to static reverse-code generation, essentially due to the inherent limitations of static analysis. 
Indeed, the state-of-the-art method of Akgul and Mooney~\cite{Akgul-Mooney-ACM04} does not support non-deterministic programs.

Overall, we believe that dynamic reverse-code generation can widen the applicability of reverse code to non-deterministic programs, while keeping memory-efficiency of reverse-code-based backtracking. 
To support our belief, we present a case study of various backtracking methods, including dynamic reverse-code generation (Section~\ref{sec:ex}).
In particular, we illustrate the usefulness of dynamic reverse-code generation by (1) considering the standard example of a multi-threaded program managing a bounded buffer and (2) measuring its memory usage for each existing backtracking method and for dynamic reverse-code generation.
As will be shown, our dynamic reverse-code generation method consumes the least amount of memory among all the backtracking methods considered.
Surely, memory efficiency, while important, is not the only factor that affects the usefulness of a backtracking method. 
We consider in Section~\ref{sec:discussion} other factors that deserve consideration such as the response time of backtracking and memory overhead of storing reverse code.

\section{Backtracking Methods} \label{sec:bm}

Four kinds of backtracking methods are found in the literature: state saving, checkpointing, static reverse-code generation and dynamic reverse-code generation.
Notably, the method for \emph{replay}~\cite{Choi-Srinivasan-SPDT98,Leblanc-Crummey-IEEE87,Russinovich-Cogswell-PLDI96}, that guarantees to reproduce the same program execution as in the past, is excluded from the list.
It is infeasible to backtrack with the replay method alone since a given program must be restarted from the beginning of its execution.

\subsection{State saving}
This method saves previous program points and previous data values while running a program forwards and restores them in a LIFO (Last In, First Out) manner when backtracking~\cite{Agrawal-DeMillo-Spafford-SPE93,Balzer-AFIPS69,Cook-CJ02,Zelkowitz-PhD71}.
The naive approach is to save a whole state-vector consisting of global variables, local variables and program point, whenever the current state of a program changes (\emph{basic state saving}). 
Alternatively, one can save only modified values (\emph{incremental state saving}).  State saving is the easiest way to perform backtracking. However its clear drawback is that memory is consumed for every statement that changes the state of a program, \eg an assignment.
Therefore, memory consumed for backtracking is directly proportional to the length of the execution path.

\subsection{Checkpointing}
This method saves previous program points and previous data values only at predefined \emph{checkpoints} (hence \squote{checkpointing}), not on every statement. 
When backtracking, one restores the state saved at the previous checkpoint (thus the program point goes back to the previous checkpoint), and runs the program forward to the desired previous program point~\cite{Agrawal-DeMillo-Spafford-IEEE91,Boothe-PLDI00}.
As in the case of state saving, it is more economical to only save changes between adjacent checkpoints than to save a state-vector at a checkpoint. 
Checkpointing can consume less memory than state saving. 
When the same variable is modified several times between adjacent checkpoints, only the variable value at the previous checkpoint is saved.
However, checkpointing is in a broad sense still a state saving because states are saved periodically. 
Therefore, memory consumed for backtracking is still directly proportional to the length of the execution path.

\subsection{Reverse-code generation}
This method first generates reverse code.
Executing this reverse code restores the previous states of a program, and as a result, a program backtracks.
As a simple example, the reverse code of \m{x:=x+1} is \m{x:=x-1}.
Such simple reverse code for self-defined assignments (i.e., assignments where the same variable is used in both sides of the assignments) are used in various reverse execution methods~\cite{Biswas-Mall-SIGPLAN99,Carothers-Perumalla-Fujimoto-PADS99}.
To handle more general cases beyond self-defined assignments, the combination of the following three techniques are used by Akgul and Mooney~\cite{Akgul-Mooney-ACM04} and by the author~\cite{Lee-VD06}: the redefine technique, the extract-from-use technique, and lastly the aforementioned state-saving technique that is used only if neither of the previous two techniques can be applied.

\def\Rulesep{2ex}
\begin{boxedfig}[!t]{1.03\figwidth}
	\begin{subfig}{1.03\figwidth}			
		\begin{formal}
		\begin{align*}			
			\integer & \in \Integer
			\quad
			\var \in \Var
			\quad
			\Exp  \in \Expression
			\quad
			\Cmd \in \Command
			\\[1.5ex]
			\Exp &::= \integer \mid \var \mid \Exp + \Exp \mid \Exp - \Exp 
			\mid \Exp \times \Exp \mid \Exp / \Exp
			\mid \BoolExp
			\qquad
			\BoolExp ::= 
			\true \mid \false \mid
			\Exp \pr{==} \Exp \mid \Exp \pr{>} \Exp \mid \pr{!} \BoolExp 
			\\
			\Cmd &::=
			  \var \pr{:=} \Exp \mid \skipc
			  \mid \Cmd;\Cmd \mid \ifstmt{\BoolExp}{\Cmd}{\Cmd}  
			  \mid \whilestmt{\BoolExp}{\Cmd}			  			  
		\end{align*}		
		\end{formal}
\vspace{8pt}	
\caption{Our minimal programming language}
\label{fig:prog}		
\end{subfig}
\hspace{3pt}\figrule		
\begin{subfig}{1.03\figwidth}			
\begin{mydisplaymath}
\xymatrix@C=27pt@R=20pt{
\stt{g:=d+1} \ar[r]
&
\stt{e:=g \times 2} \ar[r]
&
\stt{g:=e-1}
}
\vspace{-5pt}
\end{mydisplaymath}
\vspace{-1pt}
\addtocounter{figure}{-1}	
\addtocounter{subfigure}{1}	
\caption{An example execution path}
\label{fig:simple-example-rd}		
\end{subfig}
\hspace{3pt}\figrule	
\begin{subfig}{1.03\figwidth}			
\begin{mydisplaymath}
\xymatrix@C=27pt@R=20pt{
\stt{g:=d+1} \ar[r]
&
\stt{e:=g \times 2} \ar[r]
&
\stt{g:=e-1} \ar[r]
&
\stt{d:=g \times 3}
}
\vspace{-5pt}
\end{mydisplaymath}
\vspace{-1pt}
\addtocounter{figure}{-1}	
\caption{An example execution path}
\label{fig:simple-example}		
\end{subfig}
\hspace{3pt}\figrule	
\vspace{-4pt}
\addtocounter{figure}{-1}	
\caption{Our programming language and example execution paths}	
\vspace{4pt}
\end{boxedfig}

\subsubsection{Reverse-code generation techniques}

\signpost{Notations}
To explain reverse-code generation techniques, we assume a minimal imperative programming language shown in \figurename~\ref{fig:prog}.
We also use several notations. 
We refer to an execution path as \m{\epath}.
An execution path consists of assignment commands, and we refer to the assignment command at \m{n}-th position as \m{\epath(n)}.
Lastly, we refer to the left-hand-side variable of an assignment \m{\Cmd} as \m{\lhs(\Cmd)}, and the set of variables appearing in right-hand side of \m{\Cmd} as \m{\rhs(\Cmd)}.

Now, consider an \m{n}-length execution path \m{\epath} whose last assignment command (i.e., \m{\epath(n)}) is about to be reverse-executed.
Then, each of the technique works as follows.

\signpost{(1) The redefine technique}
This technique looks for the reaching definition of \m{\lhs(\epath(n))} in \m{\epath}, and re-execute that reaching definition \m{rd} if no variable in \m{\rhs(rd)} is modified between \m{rd} and \m{\epath(n)}.
As a result, a variable equivalent to the \m{\lhs(\epath(n))} reverts to its previous value.
Reverse code, in this case, is the reaching definition \m{rd}.
Meanwhile, if some variables in \m{\rhs(rd)} are modified between \m{rd} and \m{\epath(n)}, then those modified variables should be restored beforehand through the three restoration techniques of reverse-code generation.
Reverse code, in this case, is a sequence of (i) reverse code for those modified variables and (ii) \m{rd}.

As an example, consider the execution path shown in \figurename~\ref{fig:simple-example-rd}.
Note that in the end of the path the value of variable \stt{g} is modified.
The reaching definition of the last command is \stt{g:=d+1}, and variable \stt{d} is not modified behind that reaching definition. 
Thus, reverse code to reverse-execute the last command is simply \stt{g:=d+1}.

\signpost{(2) The extract-from-use technique}
Consider an execution path that ends with \m{\var_2:=\var_1+1;} \m{\var_1:=0}.
To reverse-execute the last assignment and restore the previous value of \m{\var_1}, one can execute reverse code \m{\var_1 := \var_2 - 1} that is obtained from the previous assignment, \m{\var_2:=\var_1+1}.

More generally, this technique looks for an assignment command \m{\Cmd} where \m{\rhs(\Cmd)} contains \m{\lhs(\epath(n))}.
In addition, such \m{\Cmd} must be located behind the reaching definition of \m{\lhs(\epath(n))}. 
If there is no such \m{\Cmd}, this technique cannot be used.
Suppose \m{\var_2 := f(\var_1)} is the \m{\Cmd} that is found, and function \m{f} is invertible.
If \m{\var_2} is not modified between \m{\Cmd} and \m{\epath(n)}, then \m{\lhs(\epath(n))}, which is \m{\var_1} in this case, can be restored by executing reverse code \m{\var_1 := f^{-1}(\var_2)}.
Meanwhile, if \m{\var_2} is modified, the above reverse code should be preceded by additional reverse code to restore \m{\var_2}.

The above function \m{f} does not have to involve only one variable. 
An expression like \m{\var_1 + \var_2} can be viewed as \m{f(\var_1)} by replacing \m{\var_2} with its value at that program point.
If this is the case, a use of \m{f^{-1}} in reverse code should be preceded by code that makes sure the value of \m{\var_2} is restored beforehand.

As an example, consider the execution path in \figurename~\ref{fig:simple-example} that in the end modifies the value of variable \stt{d}.
Since \stt{d} is used in the right-hand side of the first statement (i.e., \stt{g:=d+1}) before \stt{d} is re-defined at the end of the execution path, the first candidate of the reverse code for \stt{d} is \stt{d:=g-1}. 
However, \stt{g} is redefined after the first statement by \stt{g:=e-1}, so we need to recover \stt{g} beforehand. 
In this case, from \stt{e:=g$\times$2}, we get \stt{g:=e/2}. 
Putting them together, we can restore the previous value of \stt{d} through \stt{d:=e/2-1}.

Reverse code can be pre-generated before starting a debugging session.
Or it can also be generated on the fly during a debugging session.
We refer to the former as static reverse-code generation, and the latter as dynamic reverse-code generation.
We describe the differences between them in the following two subsections.

\signpost{(3) The state-saving technique}
As mentioned, if neither of the previous two techniques can be applied, then state saving is used as the last resort. 
As an example, suppose that \m{\lhs(\epath(n))} is variable \m{\var_1}, and thus the previous value of \m{\var_1} needs to be restored to backtrack.
If \m{\var_1} is assigned its previous value through user input, the redefine technique cannot be used.
Although the extract-from-use technique can be used instead in some cases, this is not always the case. 
For example, an assignment command \m{\var_2 := \var_1 \times 0} that is executed between the reaching definition of \m{\var_1} and \m{\lhs(\epath(n))} does not help with applying the extract-from-use technique because the function comprised of the right-hand side of the assignment is not invertible.

\subsubsection{Static reverse-code generation}

Akgul and Mooney employed a path-sensitive static analysis to pre-generate reverse code before starting a debugging session~\cite{Akgul-Mooney-ACM04}.
To limit the number of paths that should be considered, they unroll each loop a few times during the analysis. 
Despite that, it is occasionally possible to generate a reverse loop (i.e., a loop whose execution reverse-executes the original loop of a source program) consisting of non-state-saving code.
This is particularly true when iterations of a loop form a regular pattern such as 1, 3, 5, \m{\ldots} for an integer variable. 

To distinguish different paths in reverse code, conditionals are used. 
As an example, suppose that the reaching definition of \m{\var_1} is \m{rd_1} if a boolean expression \m{\BoolExp} holds, and \m{rd_2} if \m{! \BoolExp} holds. 
Then, possible reverse code is \m{\ifstmt{\BoolExp}{\Cmd_1; rd_1}{\Cmd_2;rd_2}}, assuming that (i) the variables used in \m{\BoolExp} are properly restored beforehand through a preceding reverse code fragment, and (ii) \m{\Cmd_1} and \m{\Cmd_2} restore the values of the variable in \m{\rhs(rd_1)} and \m{\rhs(rd_2)}, respectively.
The boolean expressions used in conditionals of reverse code are also obtained through static analysis of the source program.

While by using reverse code one can avoid saving every value change, there is a memory overhead associated with storing reverse code.
In the case of static reverse-code generation, the size of reverse code is pre-defined before starting a debugging session.
In general, this memory overhead for reverse code is offset as program execution goes on because less memory is used for state saving. 

However, non-determinism caused by multi-threading changes the dynamics. 
There are too many execution paths to consider due to interleaving between threads.
It is difficult to pre-generate reverse code of reasonable size that works for those many execution paths.
In addition, it is difficult to make static reverse code to infer dynamic non-deterministic choices (i.e., context switches between threads), if not impossible. 
Indeed, Akgul and Mooney~\cite{Akgul-Mooney-ACM04} considered only deterministic programs.

\subsubsection{Dynamic reverse-code generation}
We introduced in the author's previous work~\cite{Lee-VD06} a method that generates reverse code on the fly during a debugging session. 
This method applies the three techniques of reverse-code generation directly to the current execution path. 
Since an execution path to consider is fixed, it is tractable to generate reverse code through the three reverse-code generation techniques we earlier explained.

It is our thesis that this dynamic method is useful for backtracking in non-deterministic programs. In the next section, we substantiate this thesis by comparing the memory consumed by the backtracking methods described earlier.
In the end, we will show that our dynamic reverse-code generation consumes the least amount of memory.

\lstset{
	language=java,
	basicstyle=\ttfamily\normalsize,
	commentstyle=\ttfamily\normalsize,
        numbers=left,
        tabsize=2,
        xleftmargin=28pt,
	basewidth=0.47em
}
\begin{boxedfig}[t!]{1.03\figwidth}
\vspace{7pt}
\begin{subfig}{.5\figwidth}
\begin{lstlisting}
/***************************
  * Bounded-Buffer Example *
  **************************/
int buf[M]; // buffer size M
int g:=0; // experimental code
int empty:=M;
int full:=0;

thread Producer {
	int src[N]; // source array
	int p:=0;
	int rear:=0;
	int d:=0; // experimental code
	while (p < N) {
		wait(empty); 
		buf[rear]:=src[p];
		p:=p+1;
		rear:=rear+1;
		rear:=rear % N;
		signal(full);
		g:=d+1; // experimental code
		d:=g$\times$3;	// experimental code 
	}
}
\end{lstlisting}
\end{subfig}
\begin{subfig}{.5\figwidth}
\lstset{firstnumber=25}
\begin{lstlisting}
thread Consumer {
	int dst[N]; // destination array
	int c:=0;
	int front:=0;
	int e:=0;
	while (c < N) {
		wait(full); 
		dst[c]:=buf[front]+1;
		c:=c+1;
		front:=front+1;
		front:=front % N;
		signal(empty);
		e:=g$\times$2;	// experimental code
		g:=e-1;	// experimental code 
	}
}
// Two semaphore procedures
procedure wait(int s) {
	// atomic action
	$\langle$await(s$>$0); s:=s-1;$\rangle$ 
}
procedure signal(int s) {
	$\langle$s:=s+1;$\rangle$ // atomic action
}
\end{lstlisting}
\end{subfig}
\figrule
\vspace{-5pt}
\caption{The bounded-buffer program.}
\label{fig:bb}
\vspace{4pt}
\end{boxedfig}

\vspace{-6pt}
\section{The Case of a Bounded Buffer} \label{sec:ex}

In this section, we introduce a bounded-buffer program for which we measure the memory usage for each of the backtracking methods presented in the previous section.
Figure~\ref{fig:bb} shows a {Java-like} program manipulating a bounded buffer shared between multiple threads. 
This bounded-buffer program consists of producers, consumers and a finite buffer shared between producers and consumers. Producers put data into the buffer unless the buffer is full,
and the data in the buffer is fed to consumers unless the buffer is empty. For the sake of brevity, we consider the simple case where only one pair of producer and consumer exists. We assume that the size of the buffer is \stt{M}. Elements in the source array (denoted by \stt{src[N]}) of the producer are copied to the destination array (denoted by \stt{dst[N]}) of the consumer in order after incrementing the value by one.
For example, if \stt{src[N]} contains $\{10,20,30\}$ in the beginning, then in the end, \stt{dst[N]} is filled with $\{11,21,31\}$.
Semaphores are used to control concurrency. Two semaphore procedures \stt{wait} and \stt{signal} (assuming call-by-reference) are shown in Lines 41-48 of Figure~\ref{fig:bb}.
In the definition of \stt{wait}, command \stt{await(s>0)} blocks the execution if condition \stt{s>0} does not hold.
Therefore, if \stt{s} becomes zero indicating that the buffer is full, then the producer thread waits until the contents of the buffer are consumed by the consumer thread, and \stt{s} becomes positive as a result.
Once \stt{await(s>0)} is passed successfully, the next command, \stt{s:=s-1}, is immediately executed without being interfered by another thread as usual in semaphore.

To ensure the linearity of execution paths, we simply assume that threads are interleaved during a debugging session.
Note that this does not mean that a debugger that supports backtracking should not allow parallel (i.e. non-interleaved) execution of threads.
The linearity of execution paths can still be ensured by employing a more sophisticated method, e.g., logging the orders of access to unprotected regions  of code (i.e. regions that can cause a data race).

When interleaving threads, we assume the assignment-command-level granularity, not the usual instruction-level granularity.
In other words, interleaving cannot take place while an assignment command, e.g., \stt{p:=p+1}, is being executed. 
(Meanwhile, when the instruction-level granularity is used, an instruction for a read-access to \stt{p} and an instruction for a write access to \stt{p} can be interspersed with other instructions executed by other threads.)
Such simplification is merely to avoid excessive complications in this case study.
All the backtracking methods explained earlier can be used at both the command level and the instruction level.
For example, Akgul and Mooney perform their analysis for static reverse-code generation at the instruction level~\cite{Akgul-Mooney-ACM04}.

In order to run a program backwards, it is necessary to restore the previous program points and data values. 
First, to restore the previous program points in a non-deterministic program, it is necessary to remember what choice was made (\eg which thread became active) and when that choice was made (\eg when the thread context was switched).
Such information can be inferred by looking at the following logs: 
the order of (1) the accesses to the entries of basic blocks, and (2) program points where the thread context is switched.
We assume that those logs are recorded in every backtracking method.



Meanwhile, to restore the data values, each backtracking method presented in the previous section induces a different memory-usage pattern.
In the following, we show how many integer memory units (we assume that one integer value costs $I$ bytes) are consumed in each of the backtracking methods to support backtracking in our running example.

\subsection{Basic state saving}
This method saves a state vector whenever the state of a program changes. A state vector of the program consists of global variables and local variables. 
In the running example, we have nine integer variables (\stt{p}, \stt{c}, \stt{front}, \stt{real}, \stt{empty}, \stt{full}, \stt{g}, \stt{d}, \stt{e}), one $M$-length integer array variable (\stt{buf}) and two $N$-length integer array variables (\stt{src,dst}).
The state of a program changes eight (8) times in the loop body of \stt{Producer} and \stt{Consumer} respectively. 
Until the program terminates, \stt{Producer} and \stt{Consumer} respectively execute their loop body $N$ times.  So overall, basic state saving costs $8(9+M+2N)I\times N \times 2$ units of memory.

\subsection{Incremental state saving}
This method saves only modified values instead of state-vectors. As mentioned above, the state of a program changes eight times in the loop body of \stt{Producer} and \stt{Consumer} respectively. 
All the modified values (including \stt{buf[rear]} and \stt{dst[c]}) take integer space. 
The loop body of each thread iterates $N$ times, as pointed out previously.
Therefore incremental state saving costs $8I \times N \times 2$ units of memory, which is certainly less than the memory usage in basic state saving.


\subsection{Checkpointing}
This method saves states periodically at predefined checkpoints. The memory requirement typically depends on how densely checkpoints are set. The more coarsely they are set, the less memory is likely to be consumed but the more runtime is required to reach the previous program point from the nearest checkpoint.
In our example, let us suppose that two checkpoints are set at Lines 15 and 31. 

Since state vectors are expensive, the incremental-state-saving idea is also often used in checkpointing; only data values modified during the previous period are saved. For example, if Lines 15-20 are executed before reaching the checkpoint at Line 31, \stt{buf[rear],p,rear} and \stt{full} are saved. 

Note that \stt{rear} is saved only once. In incremental state saving, \stt{rear} is saved twice (one after Line 18, the other after Line 19).
Incremental checkpointing consumes less memory than incremental state saving when the same variable is modified several times between adjacent checkpoints.
In our example \stt{rear} and \stt{front} are modified twice, hence we can save $2I$ units of memory.
If Lines 21 and 38 are executed consecutively so that \stt{g} is modified twice in a row, we can also economize one more unit of memory.
In this optimal case, checkpointing costs $(16-3)I\times N$ units of memory, which is less than the memory usage in incremental state saving.

\subsection{Static reverse-code generation}
This method performs backtracking by running pre-generated reverse code. 
We earlier pointed out in Section~\ref{sec:bm} that the current static reverse-code generation does not help much in supporting backtracking of multi-threaded programs due to its inherent non-determinism.
Only self-defined assignments can still be used to generate non-state-saving reverse code. 
Thus, in our running example, the commands that can be restored through non-state-saving reverse code are \stt{p:=p+1} at line 17,  \stt{rear:=rear+1} at line 18, \stt{c:=c+1} at line 33, \stt{front:=front+1} at line 34, and \stt{wait}/\stt{signal} commands.
The remaining eight commands (four in \stt{Producer} and another four in \stt{Consumer}) are state-saved. 
Since the loop body of each thread iterates $N$ times, overall, the static reverse-code generation costs $4I \times N \times 2$ units of memory, which is less than the memory usage in checkpointing.

\subsection{Dynamic reverse-code generation}

This method performs backtracking by running reverse code generated on the fly.
We earlier mentioned that reverse code can be generated from the current execution path.
Recall that to restore the previous program points, we log the order of accessed basic-block entries and context-switch points.
It is possible to obtain the current execution path consisting of executed commands from that logged information and the program source code.

Different reverse code is dynamically generated depending on the current execution path.
For example, the following shows four possible scenarios among many that modifies variable \stt{d} at line 22.

\begin{boxedfig}[h!]{1.03\figwidth}
\vspace{10pt}
\begin{mydisplaymath}
\xymatrix@C=27pt@R=3pt{
\text{(a)}
&
21: \stt{g:=d+1} \ar[r]
&
22:\stt{d:=g \times 3}
&
&
\\
\text{(b)}
&
21: \stt{g:=d+1} \ar[r]
&
37:\stt{e:=g \times 2} \ar[r]
&
38:\stt{g:=e-1} \ar[r]
&
22:\stt{d:=g \times 3}
\\
\text{(c)}
&
21: \stt{g:=d+1} \ar[r]
&
37:\stt{e:=g \times 2} \ar[r]
&
22:\stt{d:=g \times 3}
&
\\
\text{(d)}
&
21: \stt{g:=d+1} \ar[r]
&
38:\stt{g:=e-1} \ar[r]
&
22:\stt{d:=g \times 3}
&
}
\vspace{-5pt}
\end{mydisplaymath}
\vspace{8pt}
\end{boxedfig}

In case of scenario (b), the thread context switches after line 21 and after line 38. 
In all the four scenarios, line 22 can be backtracked without state saving. 
In the first scenario, dynamic reverse code, \stt{d:=g-1}, can restore the previous value of \stt{d}.
The second scenario was already considered in \figurename~\ref{fig:simple-example}.
The remaining two scenarios can also be handled similarly to generate dynamic reverse code.

We can also recover array elements similarly by dynamic reverse code.  
Consider the following scenario of an execution path (array indices are replaced with their runtime values) that in the end modifies \stt{buf[0]}.

\begin{boxedfig}[h!]{1.03\figwidth}
\vspace{-1pt}
\begin{mydisplaymath}
\xymatrix@C=15pt@R=3pt{
32: \stt{dst[0]:=buf[0]+1} \ar[r]
&
33:\stt{c:=c+1} \ar[r]
&
34:\stt{front:=front+1} \ar[r]
&
35:\stt{front:=front\%N}
\\
36: \stt{signal(empty)} \ar[r]
&
15:\stt{wait(empty)} \ar[r]
&
16:\stt{buf[0]:=src[2]}
&
}
\vspace{-6pt}
\end{mydisplaymath}
\vspace{3pt}
\end{boxedfig}

Possible reverse code in this case is \stt{buf[0]:=dst[0]-1}.
The previous value of \stt{buf[0]} can be restored by executing this reverse code.
Note that the original command at line 32 is \stt{dst[c]:=buf[front]+1}.
The fact that \stt{c} and \stt{front} had value 0 can be obtained by separate reverse code for \stt{c} and \stt{front}. 

Overall, for the above execution scenarios, we only need to save the value of \stt{rear} at line 19 and the value of \stt{front} at line 35. Since the loop body of each thread iterates $N$ times, overall, dynamic reverse-code generation costs $I\times N \times 2$ units of memory, which is less than the memory usage in static reverse-code generation.
It is also noteworthy that the gap between the dynamic reverse-code generation method and the other ones becomes wider as the number of threads (i.e., \stt{Producer} and \stt{Consumer}) increases.

\section{Discussion} \label{sec:discussion}
The comparison of Section~\ref{sec:ex} is focused on the memory usage of each backtracking method.
Meanwhile, the response time of backtracking is also important. 
It is apparent that the dynamic reverse-code generation method requires more CPU cycles during a debugging session than the other methods.
Thus, it is necessary to put engineering efforts to lower response time. 
One possible way is to use both state saving and dynamic reverse-code generation altogether in a complementing way.
That is, while running a program, state saving can be used initially to guarantee a quick response time. 
At the same time, a background process can replace stored data with reverse code one by one.
Then when backtracking, reverse code is executed if it is ready; otherwise, stored data are used instead.

Another point worth considering is a memory overhead associated with storing reverse code.
The case for static reverse-code generation was earlier considered. 
The size of static reverse code is pre-defined before starting a debugging session. 
Meanwhile, the situation in dynamic reverse-code generation is more sophisticated. 
In one extreme, memory overhead can be entirely avoided if reverse code is generated only when necessary. 
As pointed out earlier, however, this will increase the response time of backtracking.
In the opposite extreme, reverse code can be continuously generated and kept in the memory as the program runs. 
While this will decrease the response time, the memory overhead in this case is proportional to the length of the current execution path.
We believe that it is wise to make a trade-off between these two extremes. 
That is, it is possible to keep in the memory only the reverse code for the last \m{n} commands of the current execution path where \m{n} is a certain trade-off number.
As an execution path gets longer, the oldest fractions of reverse code in the memory are deleted from the memory. 
Those deleted fractions of reverse code can be re-generated in the background while the latest fractions of reverse code are executed to backtrack.

\vspace{-7pt}
\section{Conclusion} \label{sec:con}

We have shown how much memory is consumed in various backtracking methods when a bounded buffer is accessed concurrently. In particular, we have pointed out why the existing static reverse-code generation does not work well on non-deterministic programs such as multi-threaded ones and why the dynamic reverse-code generation does. 
Finally, we have illustrated that our dynamic reverse-code generation can use less memory than the existing backtracking methods when applied to non-deterministic programs.

\paragraph{Acknowledgments:}
I am grateful to Dave Schmidt for his kindness. 
I admire him for his devotion to research, education, and humanity. 
I am also grateful to Olivier Danvy for encouraging me to write this paper.
Finally, my thanks also go to anonymous reviewers. 

\bibliographystyle{eptcs}
\bibliography{dave-fest}




\end{document}